\documentclass[9pt]{article}
\usepackage{spconf,amsmath,graphicx,amssymb}
\usepackage{multirow}
\usepackage{array}
\usepackage{booktabs}
\usepackage{pifont}
\newcolumntype{P}[1]{>{\centering\arraybackslash}p{#1}}
\usepackage{xcolor}

\iftrue
    \providecommand{\gz}[1]{\textcolor{black}{{#1}}}
    
    \providecommand{\yy}[1]{\textcolor{black}{{#1}}}

\else
    \providecommand{\gz}[1]{{#1}}
    
    \providecommand{\yy}[1]{{#1}}

\fi

\title{Transcription free filler word detection with Neural semi-CRFs}
\name{Ge Zhu$^1$, Yujia Yan$^1$, Juan-Pablo Caceres$^2$ and Zhiyao Duan$^1$\thanks{This work is partially supported by a New York State Center of Excellence in Data Science award.}}
\address{$^1$University of Rochester, $^2$Adobe Research \\
         \small{\tt \(  \) \{ge.zhu,zhiyao.duan,yujia.yan\}@rochester.edu \text{}  caceres@adobe.com} }
\begin{document}
\maketitle
\begin{abstract}
Non-linguistic filler words, such as ``uh" or ``um'', are prevalent in spontaneous speech and serve as indicators for expressing hesitation or uncertainty. 
Previous works for detecting certain non-linguistic filler words are highly dependent on transcriptions from a well-established commercial automatic speech recognition (ASR) system.
However, certain ASR systems are not universally accessible from many aspects, e.g., budget, target languages, and computational power. 
In this work, we investigate filler word detection system\footnote{Codebase: \gz{https://github.com/gzhu06/Filler-semi-CRF}} that does not depend on ASR systems.
We show that, by using the structured state space sequence model (S4) and neural \gz{semi-Markov conditional random fields (semi-CRFs)}, we achieve an absolute F1 improvement of $6.4\%$ (segment level) and $3.1\%$ (event level) on the PodcastFillers dataset. 
We also conduct a qualitative analysis on the detected results to analyze the limitations of our proposed system.

\end{abstract}

\begin{keywords}
filler word detection, speech disfluency, neural semi-CRFs
\end{keywords}

\vspace{-10pt}
\section{Introduction}
\label{sec:intro}
Speech disfluencies, including repetitions, word fragments, repairs and filler words, are prevalent in spontaneous speech.
Content creators often seek to remove these unwanted contents, filler words in particular. Automatically detecting filler words can speed up this laborious process.
Such system is also helpful with other speech analysis tasks. For example,
in deception detection~\cite{arciuli2010can,villar2012use}, it is shown that the instances of ``um" occur less frequently and appear to be shorter during lying~\cite{arciuli2010can}. 
A preliminary investigation~\cite{yuan2020disfluencies} finds that people with Alzheimer’s Disease (AD) use more ``uh" but less ``um". Filler word detection may also be helpful for automatic speech transcription (ASR), as it is shown that filler words can lead to an increase in ASR error rates~\cite{goldwater2008words,horii2022end}.

Due to the broad interest in detecting filler words, many studies have been proposed in recent years.
In natural language processing (NLP), disfluency correction of spontaneous speech transcription is heavily investigated for downstream language processing tasks~\cite{bach2019noisy,wang2020multi,jamshid-lou-etal-2018-disfluency}. 
However, obtaining an accurate transcription that includes filler words from speech directly is a nontrivial task; it requires either human effort or a sophisticated verbatim ASR system developed on a large annotated dataset.
There are also studies on filler word detection on speech signals by integrating with ASR systems.
Inaguma et al.~\cite{inaguma2018end} proposed to integrate event detection modules to simultaneously recognize speech and detect and fillers. 
All these above-mentioned systems depend on a well-performing ASR system, which may be inaccessible for applications with limited budget and computational power~\cite{rocholl21_interspeech}. In addition, many applications requires the filler words detection to be performed on-device. Therefore, it would be desirable to develop automatic filler word detection methods that do not require verbatim transcription.


Filler detection based on low-level acoustic features~\cite{audhkhasi2009formant} have been built using Hidden Markov Models~\cite{salamin2013automatic}, Conditional Random Fields~\cite{schuller2008static} and neural networks~\cite{kaushik2015laughter,gupta2013paralinguistic}.
In previous neural network based systems, frame-level classifiers are first trained using convolutional neural networks (CNN), and then the raw probability outputs are fed into a separate temporal post-processor, which involves careful design and tuning. 
Due to the lack of public datasets, previous works were mainly developed on well-controlled small scale datasets. 
Moreover, they only evaluate their systems at the frame level without reporting performance at the event level~\cite{zhu22e_interspeech}. 
The \textit{event-based} metrics~\cite{mesaros2016metrics} measure the ability to detect the correct holistic event at the correct temporal position. 
The performance of the event-based metrics is usually much lower than \textit{segment-based} metrics. 
We believe that event-based metrics are more suitable for measuring the performance in practice because of its discrete nature.

In this paper, we propose a system that improves transcription free filler word detection:  
we incorporate a recently proposed lightweight S4 backbone~\cite{gu2021efficiently} and shows that it improves the overall detection performance;
we further use neural semi-CRFs adapted for temporal event detection~\cite{yan2021skipping} as a direct formulation for the final output that avoids manually designed post processing steps. 

\begin{figure}[!t]
\centering
\includegraphics[width=0.6\columnwidth]{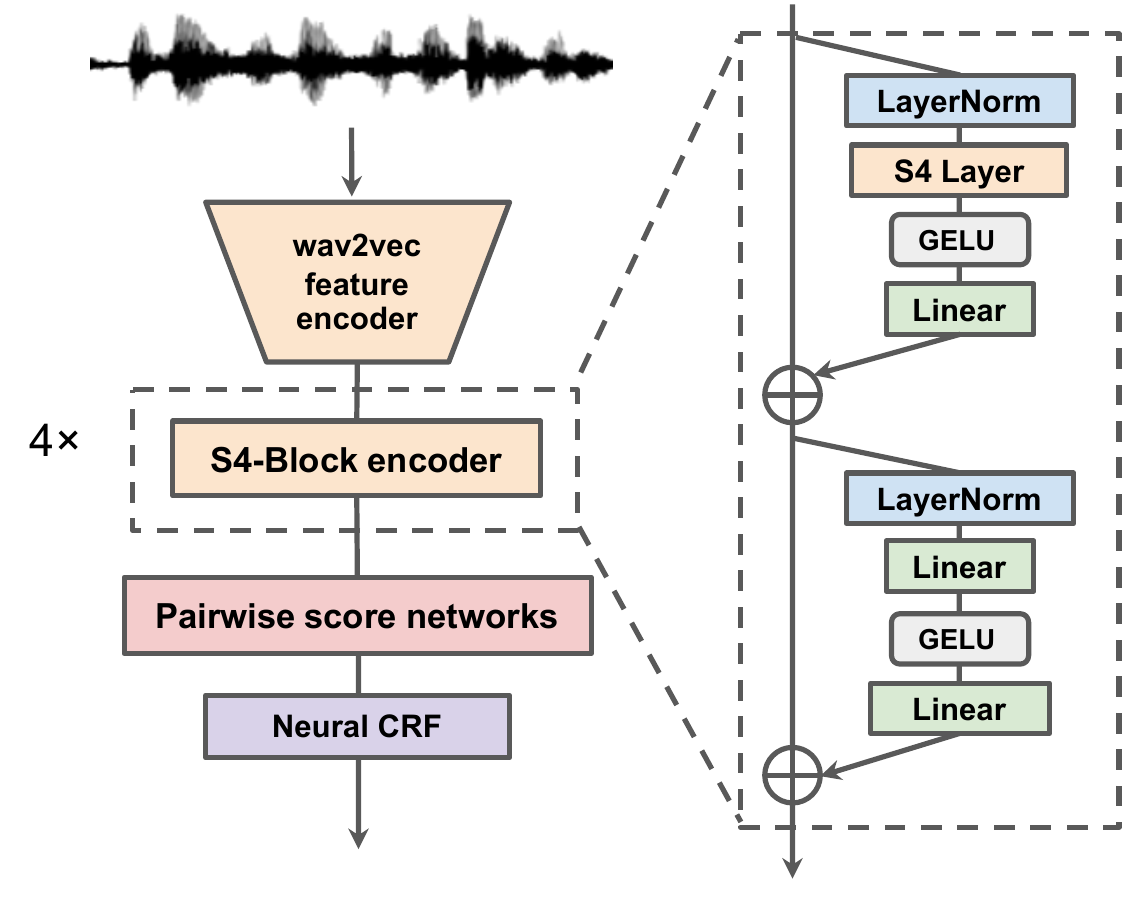}
\caption{Proposed transcription-free filler word detector.}
\label{fig:sd}
\end{figure}

\section{Proposed system}
In the problem of filler word detection from speech signals, we are given $n$ frames of input features $x = (x_1, ... , x_n)$, such as mel-frequency cepstral coefficients (MFCC), the target is to output the intervals $\{(u_i, v_i)\}$ for the filler events, where $u_i$ and $v_i$ indicate the starting time and ending time, respectively. 
In our proposed system shown in Fig.~\ref{fig:sd}, we feed wav2vec encoded feature maps into a context encoder, followed by a pairwise frame index score network. 
Eventually, the semi-CRF layer directly outputs the target event time interval. 

\textbf{S4 encoder.}
During classification, partial word frames with similar phonemes as `um' or `uh' can be misclassified without referring to neighboring contexts. 
One typical example could be the leading part of `um-brella'.
A network architecture that is able to model temporal characteristics is crucial to avoid such false positives. 
In VC-FillerNet~\cite{zhu22e_interspeech}, convolutional recurrent neural networks (CRNN) is used as the classifier backbone, which is widely used in sound event detection (SED). 
In CRNN, the stacked convolutional layers on the top act as feature extractors to learn discriminative time-frequency features. 
The recurrent layers integrate the extracted features over time to model the context information. 

We propose to apply the structured state space sequence (S4) model~\cite{gu2021efficiently} to replace the CRNN backbone for a fast and accurate model and potentially for real-time applications. 
S4 is a special case of the discretized state space model (SSM). 
The discretized state space model inherits the properties of both CNNs and RNNs~\cite{gu2021combining}: 
\begin{align}
    h_k=\bar{A}h_{k-1}+\bar{B}x_{k},~~~~y_k=\bar{C}h_{k}+\bar{D}x_{k},
    \label{eq:dss}
\end{align}
where $\bar{A},\bar{B},\bar{C},\bar{D}$ are discretized SSM parameters. 
We can rewrite Eq.~\eqref{eq:dss} in the form of a standard RNN setting $\bar{D}=0$:
\begin{align}
    h_k=f(h_{k-1},x_{k}),~~~~y_k=g(h_{k}),
    \label{eq:rnn}
\end{align}
where $f$ and $g$ are linear transforms. 
We can also rewrite Eq.~\eqref{eq:dss} in the form of convolution mimicking CNN by unrolling all of the states defined in Eq.~\ref{eq:dss}, keeping $\bar{D}=0$:
\begin{align}
    y=\bar{K}*x,~~~~\bar{K}=(\overline{CB},\overline{CAB},...,\overline{CA}^{n-1}\overline{B}),
    \label{eq:cnn}
\end{align}
where $\bar{K}$ can be viewed as the convolutional kernel. 
Unlike vanilla CNNs that commonly utilize small kernels and deep layers, the size of $\bar{K}$ is set with length of the entire sequence, achieving a long receptive field with only one layer.

\begin{figure}[!t]
\centering
\includegraphics[width=0.7\columnwidth]{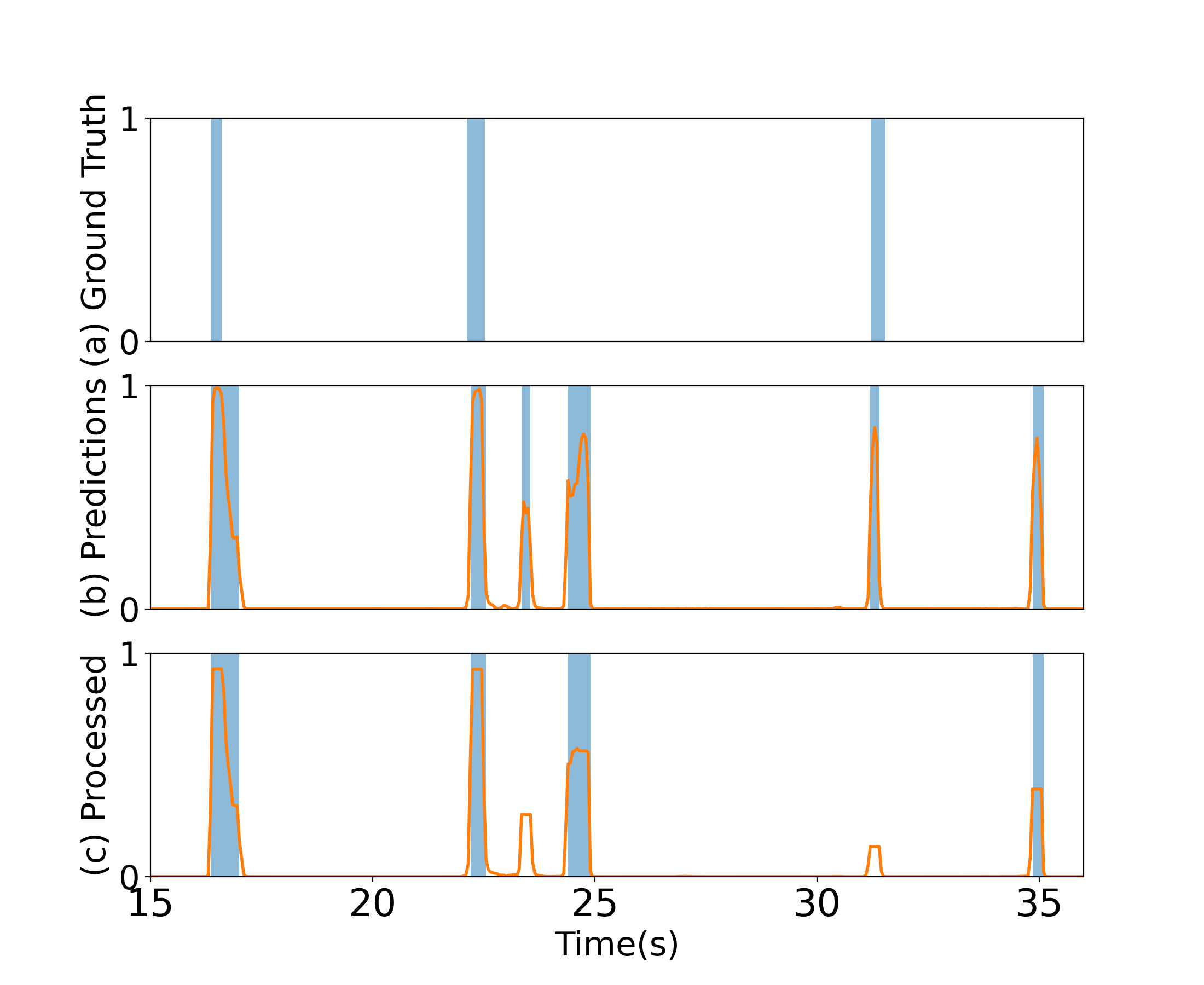}
\caption{Examples of the post-processing pipeline of frame-level filler prediction outputs. Top row: blue columns are ground-truth fillers. Middle row: the orange curve is the filler prediction confidence from VC-FillerNet and the blue columns are detected fillers. Bottom row: median filtered confidence curve and resulting fillers.}
\label{fig:pp}
\end{figure}

Specifically, S4 applies a special parameterization of $\bar{A}$ using a diagonal plus low-rank matrix for the SSM, which has two key properties: First, this structured representation is small in size and allows the faster computation of the convolution kernel $\bar{K}$. 
Second, it allows the SSM to capture long-range sequence information and achieves strong results in sequence classification~\cite{gu2021efficiently}. To overcome the limited representation ability induced by the linear transform in Eq.~(\ref{eq:rnn}), we employ the S4 block~\cite{goel2022s} with an additional feed-forward network using the pre-layernorm design~\cite{xiong2020layer}, shown in the right part of Fig.~\ref{fig:sd}. 
The advantages of the proposed backbone is that it achieves a large receptive field over the entire audio feature sequence with a smaller amount of parameters. 

\textbf{Semi-CRF layer.} 
Frame-level filler detectors, which output the framewise likelihood of a position being a filler, usually rely on ad hoc post processing pipelines.
Those post-processing steps are usually heuristics-based, typically including smoothing, thresholding, etc, for transforming noisy framewise estimations into clean event intervals. This process is shown in Fig.~\ref{fig:pp}, in this example, we manually tune the median filtering size to 9 to filter the raw predictions for fillers from VC-FillerNet~\cite{zhu22e_interspeech}; we can see that true and false positives are filtered out.

Semi-CRFs~\cite{sarawagi2004semi} was originally proposed to address sequence tagging problems in NLP, and recently it was successfully applied to distinguishing disfluent chunks for transcriptions~\cite{ferguson2015disfluency} and acoustic modeling~\cite{lu2016segmental}. 
Similarly, we can use semi-CRFs to output the filler word intervals from a sequence of features from the encoder network. 
Previous approaches using semi-CRFs are meant to find a segmentation of input sequences, of which each segment is labeled.
For the scenario of finding disfluent chunks from audio frames, of which positions are discretized from continuous positions, we incorporate a specially adapted version of neural semi-CRFs as proposed in \cite{yan2021skipping}. 
The difference between this variant and the original versions are as follows.
The original semi-CRF models the conditional probability over all possible segmentations; while the one in \cite{yan2021skipping} is over all possible sets of non-overlapping intervals, \yy{which contains no specific segment for non-events}.
Moreover, intervals are allowed to overlap on endpoints (onset/offset) in \cite{yan2021skipping} for allowing an offset and onset to be discretized in the same frame.

Each set of intervals for a specific event type $e$ is modeled as the following conditional probability given the input $x$: 
\begin{equation}
    \log p_{\theta}(Y_e|x)=\log\sum_{(i,j,e)\in Y_e} \exp(f(i,j, e)) - \log(Z(e)),
\end{equation}
where $Y_e$ denotes the set of events of a specific event type $e$ that are non-overlapping except for the endpoints, and $Z(e)$ is the normalization factor. 
The function $f(i, j, e)$ assigns a score to an interval $[i, j]$ that is labeled as the event type $e$. 
For simplicity we omit the skip score in \cite{yan2021skipping}.

We follow the same scoring module design as \cite{yan2021skipping}, which is a three-layer feed-forward neural network followed by a three-layer 2-d CNNs that outputs a score tensor with shape $T_{1}\times T_{0}\times N$. 
Here $T_{1}$ and $T_{0}$ are the ending and beginning positions of the entire sequence respectively, and $N$ corresponds to the number of event types.
\yy{The final training objective is to maximize the total conditional log-likelihood of target event types}:
\begin{equation}
    \log p_{\theta}(Y|x)=\sum_e^{N}\log p_{\theta}(Y_e|x).
\end{equation}

For inference, we use the same Viterbi algorithm as in \cite{yan2021skipping} to infer the most likely set of intervals.
More specifically, 
\begin{equation}
    Y^*_e = arg\max_{Y_e}  \log p_{\theta} (Y_e| x).
\end{equation}
We refer the readers to \cite{yan2021skipping} for more details.

\section{Experimental Design}

\textbf{Dataset.}
\label{sec:data}
We train and evaluate our proposed system on the PodcastFillers dataset~\cite{zhu22e_interspeech} that consists of 145 hours of audio from over 350 speakers in English. 
The annotations consist of over 85k manually annotated audio events with onsets and offsets including approximately 35k filler words and 50k non-filler events. 
Practically, other than the target ``Filler" class, we also include the events from categories of ``Speech", ``Laughter'', ``Breath'' and ``Music'' to train the frame-level encoder on top of the semi-CRF layer to stablize the training.
We follow the original dataset split which is based on podcast shows to avoid speaker identity overlapping. 

During training, we filtered out apparently non-voiced regions from PodcastFillers with a voice activity detector (VAD)~\cite{chen2020voice}.
This filtering procedure was performed with a small threshold to achieve a high recall. As a result, the majority of silence, music and noise were removed.

\textbf{Training details.}
Based on the fact that the actual fillers usually have variable length, ranging from 50ms to nearly 1s~\cite{kaushik2015laughter,shriberg1999phonetic}, we use the label resolution of 50ms and perform training on audio segments that are 2s long. 
As a comparison, the system proposed in \cite{zhu22e_interspeech} is trained on 1s speech segments with labels at a temporal resolution of 100ms. 

To achieve better performance, we use the pretrained feature encoder from wav2vec~\cite{schneider2019wav2vec} that encodes 16khz audio into a sequence of temporal embeddings with a hop size of 10ms. 
The receptive field for this pretrained encoder network is 30ms.
To make the system deployable, we demand a small-sized model, therefore we do not use the full encoder in \cite{schneider2019wav2vec}. Instead, we only use the first few layers, because we believe that the low-level acoustic features are sufficient for detecting fillers.
We use a batch size of 64 and train for 30 epochs using AdamW optimizer with a weight decay of 1e-2 at an initial learning rate of 1e-3 using the cosine annealing scheduler. 

\textbf{Evaluation.}
We use segment-based and event-based metrics originally proposed for SED~\cite{mesaros2016metrics} to evaluate the filler word detection performance. 
Segment-based metrics compare the system outputs and the ground truth on a fixed time grid.
Event-based metrics directly compare the estimated sound events with the ground-truth events by using a maximum cardinality matching between predictions and ground truths. 
A predicted event is considered a true positive if and only if it is assigned with the correct label and its onset and offset are within a deviation threshold (200 ms in this work) from the reference event's onset and offset.

\begin{table*}[t]
\caption{Segment- and event-based results ($\%$) of our proposed systems on the test split of PodcastFillers. 
All trainable parameters are single precision floating points and the numbers of parameters exclude the 5M parameters from the wav2vec feature encoder. MHA (multi-head self-attention) stands for transformer encoder.
}
\centering
\scalebox{0.75}{\begin{tabular}{P{105pt}|P{45pt}P{35pt}P{40pt}P{55pt}P{28pt}P{27pt}P{15pt}P{28pt}P{27pt}P{15pt}}
\toprule
\multirow{2}{*}{\textbf{System}}&\textbf{Label Res.}&\multirow{2}{*}{\textbf{Backbone}}&\multirow{2}{*}{\textbf{Semi-CRF}}&\multirow{2}{*}{\textbf{\# Params}}&
\multicolumn{3}{c}{\textbf{Segment-based (\%)}} &
\multicolumn{3}{c}{\textbf{Event-based (\%)}}\\ 
\cline{6-8}\cline{9-11}
&(ms)&&&&Precision&Recall&F1&Precision&Recall&F1\\
\hline
\textit{Transcription-based}\\
AVC~\cite{zhu22e_interspeech}&-&CNN&&-&\textbf{93.0}&\textbf{95.4}&\textbf{94.2}&\textbf{91.7}&	\textbf{94.0}& \textbf{92.8}	 \\ 
\hline
\textit{Transcription-free}\\
VC-100-CRNN~\cite{zhu22e_interspeech} &100&CRNN&&344k& 78.4&69.7 &73.8&74.8&76.9&73.8\\
VC-50-CRNN&50&CRNN&&  344k&79.5	& 75.4&77.4	&72.7	&77.1&	74.8\\ 
VC-50-CRNNCRF&50&CRNN&\checkmark&  357k &78.0&79.2&78.6	&75.3&74.9&75.1\\ 
VC-100-MHA&100&Transformer&&213k &72.4&65.9&69.0&68.8&66.2&67.5\\
VC-50-MHA&50&Transformer&&213k&77.4&72.7&75.0&72.4&73.1&72.7\\ 
VC-50-MHACRF&50&Transformer&\checkmark& 226k&64.1&84.2	&72.7&62.0&80.2&69.9\\ 
VC-100-S4&100&S4&&215k &76.2&75.1&75.6&75.4&74.6&75.0\\
VC-50-S4&50&S4&&215k &82.1&74.3&78.0&75.0	&76.4&75.7\\ 
\textbf{VC-50-S4CRF (Ours)}&50&S4&\checkmark& 221k&80.2&80.1&\textbf{80.2}	&79.5&74.5&\textbf{76.9}\\ 
\bottomrule
\end{tabular}}
\label{tab:results}
\vspace{-2mm}
\end{table*}

\textbf{Ablations.} 
We run ablation studies, with model configurations shown in Table~\ref{tab:results}, on the validation split of PodcastFillers to understand the impact of label resolution, model backbones and the semi-CRF layer.
We first compare the temporal label resolution at 100ms and 50ms respectively. 
For backbone comparison, we include CRNN, S4 block and encoder-only transformer with a similar model size.
When assessing effects of the semi-CRF layer, to make the ablations comparable, we add an additional median filter of size 3 for the frame-wise classifier without the semi-CRF layer. 
For the systems with the semi-CRF layer, we directly evaluate on the test split because there is no explicit hyperparameters involved for tuning in this case.

\section{Results}
\label{sec:res}

\begin{figure}[!t]
\centering
\includegraphics[width=0.95\columnwidth]{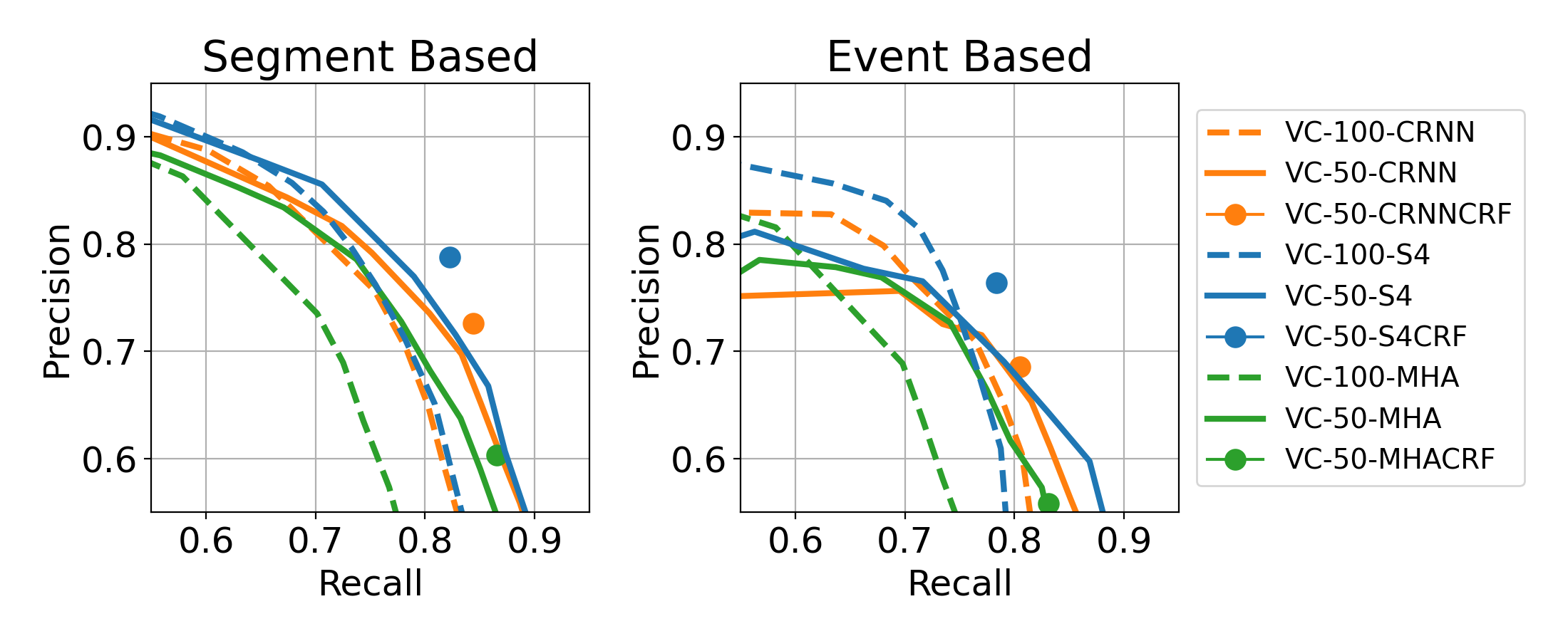}
\caption{PR curves for the ablations on the validation split of PodcastFillers. Systems with the semi-CRF layer do not encode threshold for tuning explicitly during training, therefore the result is only one value instead of a curve. }
\label{fig:valcomp}
\end{figure}

\textbf{Ablation studies.}
We begin with analyzing the influence of the label resolution by comparing the solid lines versus dashed lines on the precision-recall (PR) curves shown in Fig.~\ref{fig:valcomp}.
For the segment-based metrics, we can see that for different backbone model classes, the systems with a fine label resolution almost consistently outperforms its counterpart with a coarse label resolution. 
This improvement shows that a finer label resolution is helpful in locating the boundaries of the filler words.
For event-based metrics, finer labels tend to have higher recall but lower precision. 
Since a finer label resolution yields more predictions, therefore, it is more prone to misclassification and leads to lower precision.

In the comparison of the backbones differently colored in Fig.~\ref{fig:valcomp}, we see that when fixing the label resolution, the PR curves of S4 based systems marginally outperforms the CRNN backbone in both metrics, indicating its efficiency and effectiveness in detecting fillers. 
\gz{In S4-based models, the computation of convolution in Eq.~\eqref{eq:cnn} requires fast Fourier transform (FFT) and inverse FFT. 
As a result, they are not as efficient as CRNN-based models. 
It is shown that such convolution can be replaced with parallel scan to improve efficiency~\cite{smith2022simplified}, but this optimization is beyond the scope of this paper.}
Among all of the backbones, the transformer underperforms both CRNN and S4. 

After inserting the semi-CRF layer, we observe that it improves over the vanilla systems using S4 and CRNN backbones on both evaluation metrics without any tuning or threshold calibration. 
But the transformer variant still makes no improvement on the event-based detection. 

%

%


\textbf{Comparison to baselines.}
We compare the detection performance of VC-FillerNet variants on the test split of PodcastFillers, shown in Table~\ref{tab:results}. 
We see that VC-50-S4CRF consistently achieves the best performance in both metrics. 
Although there is still a large gap from the transcription-based system, VC-50-S4CRF achieves 6.4$\%$ absolute improvement in segment-based F1 and 3.1$\%$ in event-based F1 with a smaller model size over the VC-100-CRNN baseline. 

\textbf{Limitations.} We conduct a qualitative analysis on the whole test split to find out the mistakes that transcription-free systems tend to make during the detection. 
First of all, we track all of the false positives and misses respectively by collecting the events that failed to be matched during the event-based metrics evaluation. 
Then by looking at the false positives, we analyze the words that are easy to be misclassified as fillers; 
Similarly, we collect the surrounding words in the misses to analyze.

We observe that the major source of false positives comes from the confusion between words with similar phonemes to fillers: The top five cases are ``a", ``the", ``uh-huh", ``oh" and ``I". 
An ASR system, on the other hand, can provide more context for distinguishing those ambiguous cases with the help of language modeling and alignment.
For miss-detected fillers, the top five words that appear before the fillers are ``and", ``but", ``the", ``it's" and ``that". When listening to the audio samples, we found that the miss-detected fillers are mostly ``uh" which are generally short and occur at the utterance medially~\cite{clark2002using}. 
They mostly happened during the linking words pronunciation, for example, ``\textit{... and [uh] probably...}". 
Finally, we also find a decent amount of misaligned fillers, longer than the 200ms threshold, causing both false positive and missed errors.

\vspace{-10pt}
\section{Conclusion}
\label{sec:conclusion}
In this work, we presented a transcription free filler word detection system, which employs a S4 backbone and a neural semi-CRF layer. 
The S4 block acts as an efficient sequence encoder for the input features and semi-CRF layer directly model filler events and output intervals without any post-processing or threshold calibration.
Future work can concentrate on closing the gap from the transcription-based system and extending this framework to the detection of general speech disfluencies without transcriptions.


\small
\bibliographystyle{IEEEbib}
\bibliography{refs}

\end{document}